\documentclass[
preprint,
amsmath,amssymb,longbibliography,aps,pra]{revtex4-2}

\usepackage{graphicx}
\usepackage{dcolumn}
\usepackage{bm}
\usepackage{extarrows}

\usepackage{float}
\usepackage{verbatim}

\usepackage{xcolor}
\usepackage[pdfpagelabels,bookmarks,breaklinks]{hyperref}
\definecolor{darkblue}{RGB}{0,0,127} 
\definecolor{darkgreen}{RGB}{0,150,0}
\definecolor{mypurple}{RGB}{146,31,169}
\definecolor{mydarkblue}{RGB}{31,19,181}
\definecolor{mydarkgreen1}{RGB}{56,112,0}
\definecolor{mydarkgreen2}{RGB}{41,134,34}
\hypersetup{colorlinks, linkcolor=blue, citecolor=magenta, filecolor=red, urlcolor=blue}
\hypersetup{pdfpagemode=UseNone} 
\hypersetup{pdfstartview=FitH} 

\DeclareMathOperator{\sech}{\mbox{sech}}                   
\DeclareMathOperator{\ssech}{\mbox{\scriptsize sech}}      

\newcommand{\eref}[1]{(\ref{#1})}

\newcommand{\hop}[2]{\hat{#1}_{#2}}

\newcommand{\Revise}[1]{\textcolor{black}{#1}}      

\begin{document}

\title{Linear-optical approach to encoding qubits into harmonic-oscillator modes via quantum walks}

\author{Jun-Yi Wu$^{1,2,3}$ and Shin-Tza Wu$^4$}
\email{phystw@gmail.com}
\affiliation{
$^1$Department of Physics, Tamkang University, New Taipei 251, Taiwan \\
$^2$Center for Advanced Quantum computing, Tamkang University, New Taipei 251, Taiwan
\\
$^3$Physics Division, National Center for Theoretical Sciences, Taipei 10617, Taiwan\\
$^4$Department of Physics, National Chung Cheng University, Chiayi 621, Taiwan}

\date{\today}

\begin{abstract}
We propose a linear-optical scheme that allows encoding grid-state quantum bits (qubits) into a bosonic mode
using cat state and post-selection as sources of non-Gaussianity in the encoding.
As a linear-optical realization of the quantum-walk
encoding scheme in [Lin {\em et al.}, Quantum Info. Processing {\bf 19}, 272 (2020)], we employ the cat state
as a quantum coin that enables encoding approximate Gottesman-Kitaev-Preskill (GKP) qubits through quantum walk
of a squeezed vacuum state in phase space. We show that the conditional phase-space displacement necessary
for the encoding can be realized
through a Mach-Zehnder interferometer (MZI) assisted with ancillary cat-state input under appropriate parameter regimes.
By analyzing the fidelity of the MZI-based displacement operation, we identify the region of parameter space over
which the proposed linear-optical scheme can generate grid-state qubits with high fidelity.
With adequate parameter setting, our proposal should be accessible to current optical and
superconducting-circuit platforms
in preparing grid-state qubits for bosonic modes in the, respectively, optical and microwave domains.
\end{abstract}

\maketitle

\section{\label{sec:int}Introduction}
Computation and information processing based on quantum-mechanical systems
have open up new horizons to information sciences. Quantum algorithms and communication
protocols surpassing their classical counterparts in computational efficiency and communicational
security, respectively, have been developed based on this new paradigm \cite{Ni00}.
The physical realizations of these ideas, however, have been impeded by the extreme fragility of
quantum systems against external perturbations. One common remedy to this is to resort to
techniques of quantum error corrections, which seek to protect logical information by redundantly
encoding the information into a larger quantum mechanical state-space \cite{Li13}. For instance,
in computational schemes with an underlying two-dimensional logical basis, one may
choose to encode logical information using a manifold of two-level systems in a tailored way that can allow
detecting and correcting the predominant errors of the specific physical system. This approach, however,
tends to incur heavy resource overheads in the computation and poses challenges to its large-scale
implementation. Alternately, one may choose to adopt computational schemes that
employ quantum systems with continuous spectra, usually referred to as continuous-variable systems,
and exploit their infinite-dimensional state spaces for the encoding. In the case of
a bosonic field mode, the logical information can be encoded using eigenstates of its canonical
coordinates, which are usually compared to the position and momentum of a quantum simple harmonic oscillator
\cite{Br03,Br05}. In the presence of noises, these basis states would be exposed to
random shifting over phase space that can result in ``shift errors" in the logical information.
In order to tackle such errors, Gottesman, Kitaev, and Preskill (GKP) propose to
embed quantum bits (qubits) into the bosonic mode in a way that the logical basis states would comprise
superposition of uniformly spaced position eigenstates with period $2\sqrt{\pi}$ \cite{Go01}
\begin{eqnarray}
|0\rangle_L &\equiv&\!\!\sum_{s=-\infty}^\infty\!\!|2s\sqrt{\pi}\,\rangle_x
= \frac{1}{\sqrt{2}}\!\sum_{s=-\infty}^\infty\!\!|s\sqrt{\pi}\,\rangle_p \, ,
\nonumber \\
|1\rangle_L &\equiv& \!\!\sum_{s=-\infty}^\infty\!\!|(2s+1)\sqrt{\pi}\,\rangle_x
= \frac{1}{\sqrt{2}}\!\sum_{s=-\infty}^\infty\!\!(-1)^s |s\sqrt{\pi}\,\rangle_p ,
\label{ideal_GKP}
\end{eqnarray}
where $|x\rangle_x$ and $|p\rangle_p$ are, respectively, the position
and momentum eigenstates of the bosonic mode. In the position-space representation,
the \Revise{codeword states} $|0\rangle_L$ and $|1\rangle_L$ are superposition of
delta functions located at, respectively, even and odd multiples of
$\sqrt{\pi}$. For sufficiently small shift errors in the position quadrature, one can
thus detect and correct them through position measurements. Likewise,
the momentum-space structures of the conjugate basis states
$\{|\pm\rangle_L\equiv(|0\rangle_L\pm|1\rangle_L)/\sqrt{2}\}$ enable the
detection and correction of small shift errors in the momentum quadrature.
In the quantum-mechanical phase space, the Wigner functions of the \Revise{codeword states} \eref{ideal_GKP}
take the forms of checkerboard, square-lattice structures, which can also be extended to
more general ``grid states", such as \Revise{codeword states} with rectangular or hexagonal lattice
structures in phase space \cite{Go01}. In the present work, however, we will be focusing mainly on
the square-lattice GKP \Revise{codeword states} \eref{ideal_GKP}.

Despite the elegance of the GKP scheme, the ideal \Revise{codeword states}
\eref{ideal_GKP} are unphysical, as their preparation would require infinite
squeezing, and hence infinite energy. In practice, therefore, it is necessary to approximate
\eref{ideal_GKP} with finitely squeezed states, such as uniformly
spaced Gaussian spikes modulated by Gaussian envelopes \cite{Go01}. Experimental preparation of
such approximate codeword states, nonetheless, turn out extremely challenging due to the highly non-Gaussian
nature of these states. It is after decades of theoretical \cite{Tr02,Pi04,Va10,Et14,Te16,Mo17,We18,Ea19,Li20}
and experimental endeavors that approximate GKP \Revise{codeword states} have finally been achieved in the past few years
using trapped-ion \cite{Fl19} and superconducting-circuit platforms \cite{Ca20}. Moreover, by incorporating
(dissipative) stabilization mechanisms \cite{Ro20,Ne22} into the error-correction scheme, experimentalists
are now capable of \Revise{extending the coherence time of GKP logical qubits by a factor of $\sim 230\%$
compared with that of native qubits encoded through the lowest two oscillator levels of
the constituent cavity} \cite{Ne22,Si23}.
Notably, GKP encoded qubits have also been found to hold the key to
large-scale fault-tolerant quantum computing for state-of-the-art architectures based on
superconducting-circuit \cite{Te20} and optical \cite{Bo21} systems. In spite of this,
nevertheless, in the optical domain (approximate) GKP \Revise{codeword states} remain elusive in current optical
laboratories\Revise{, except for a recent proof-of-principle demonstration of optical GKP states
with broad and overlapping Gaussian spikes \cite{Ko24}.}
It is our goal in this paper to partially bridge this gap by offering an experimentally feasible scheme
for engineering approximate GKP codeword states in a linear-optical setting that will be accessible to
present day optical platforms. In particular, we will resort to a quantum walk (QW) approach developed by
Lin {\em et al.} in Ref. \cite{Li20} and supply a linear-optical realization of the scheme.
\Revise{Essentially, we will utilize cat states as resources together with post-selections in
building approximate GKP codeword states from input squeezed vacuum state. Notably,
our approach can be adjusted to allow for cat resource states with small amplitudes, as of those in
current optical platforms. As we will show, our encoding process consists of repeated conditional phase-space
displacements of the squeezed vacuum state enabled by the cat resource state along with post-selection, which
has a $50\%$ success rate per run.
These are in contrast to existing linear-optical schemes such as in Ref.~\cite{Va10}, where one requires
{\em squeezed} cat-state resources and has a low post-selection rate. In particular,
with the supply of (unsqueezed) cat states as in Ref.~\cite{Ko24}, when supplemented with
a squeezed vacuum state input and the ability for post-selection,
our linear-optical scheme shall be able to prepare approximate GKP codeword states with squeezed Gaussian spikes,
as we will demonstrate in our results. }

In what follows, we will begin with a brief summary of the QW encoding scheme of
Ref.~\cite{Li20}. We will then demonstrate a linear-optical scheme that is capable of
preparing approximate GKP states following the QW mechanism. Since the proposed encoding
mechanism is subject to specified conditions, the codeword states prepared would have
non-ideal fidelity. We analyze the quality of the \Revise{codeword state} through the fidelity of the
displacement operation that underlies the encoding process, and
identify the parameter regimes over which the encoding has better performance.
Finally, we summarize our findings in Sec.~\ref{sec:concl} and point out
possible future directions. For presentational clarity, \Revise{we relegate detailed calculations
to the Appendices}.

\section{\label{sec:main} Quantum-walk encoding and its linear-optical realization}
\subsection{\Revise{Quantum-walk encoding}}
In Ref.~\cite{Li20} a theoretical framework is proposed to prepare bosonic codeword states with
grid-like structures, such as the GKP states \eref{ideal_GKP}, through one-dimensional QW of
a bosonic mode in phase space. The idea is to generate the codeword structures in phase space through
conditional displacements of the bosonic mode via QW processes. Crucially, since the phase-space pattern
of a QW is controlled by its coin-toss transformation \cite{Ke03,Ma14}, the codeword structure of a
QW encoding would likewise depend essentially on the underlying coin-toss operation \cite{Li20}.
Suppose the bosonic mode has mode operator $\hat{a} = (\hat{x} + i\hat{p})/\sqrt{2}$, where
$\hat{x}$ and $\hat{p}$ are the corresponding, respectively, ``position" and ``momentum" quadrature operators.
The commutation relation $[\hat{a},\hat{a}^\dagger]=1$ for the mode operator thus yields that
$[\hat{x},\hat{p}]= i$, which amounts to setting $\hbar=1$ here. If the bosonic mode is prepared initially in
a position-squeezed state, QW of the state will generate coherent superposition of position-squeezed states subsequently.
Explicitly, in terms of the phase-space displacement operator $\hat{D}(\alpha)\equiv
\exp\{\alpha\hat{a}^\dagger-\alpha^*\hat{a}\}$ and the
squeezing operator $\hat{S}(\zeta)\equiv\exp\{-\frac{\zeta}{2}\hat{a}^{\dagger 2}+\frac{\zeta^*}{2}\hat{a}^2\}$,
we will be dealing with superposition of squeezed coherent states \cite{Ba97}
\begin{eqnarray}
|\alpha, \zeta\rangle
\equiv \hat{D}(\alpha)\, \hat{S}(\zeta) |0\rangle
\label{qx}
\end{eqnarray}
with $|0\rangle$ being the vacuum state of the bosonic mode. It should be noted that, as a consequence of having $\hbar=1$ here,
when applying the displacement operator $\hat{D}(\alpha)$ with real $\alpha$ to any position eigenstate $|x\rangle_x$
we would get
\begin{eqnarray}
\hat{D}(\alpha) |x\rangle_x = |x+\sqrt{2}\,\alpha\rangle_x \, ,
\label{sqrt2}
\end{eqnarray}
where the displacement in the position eigenstate incurs an additional factor of $\sqrt{2}$ \cite{note_sqrt2}.

\begin{figure}
\centering
\includegraphics*[width=90mm]{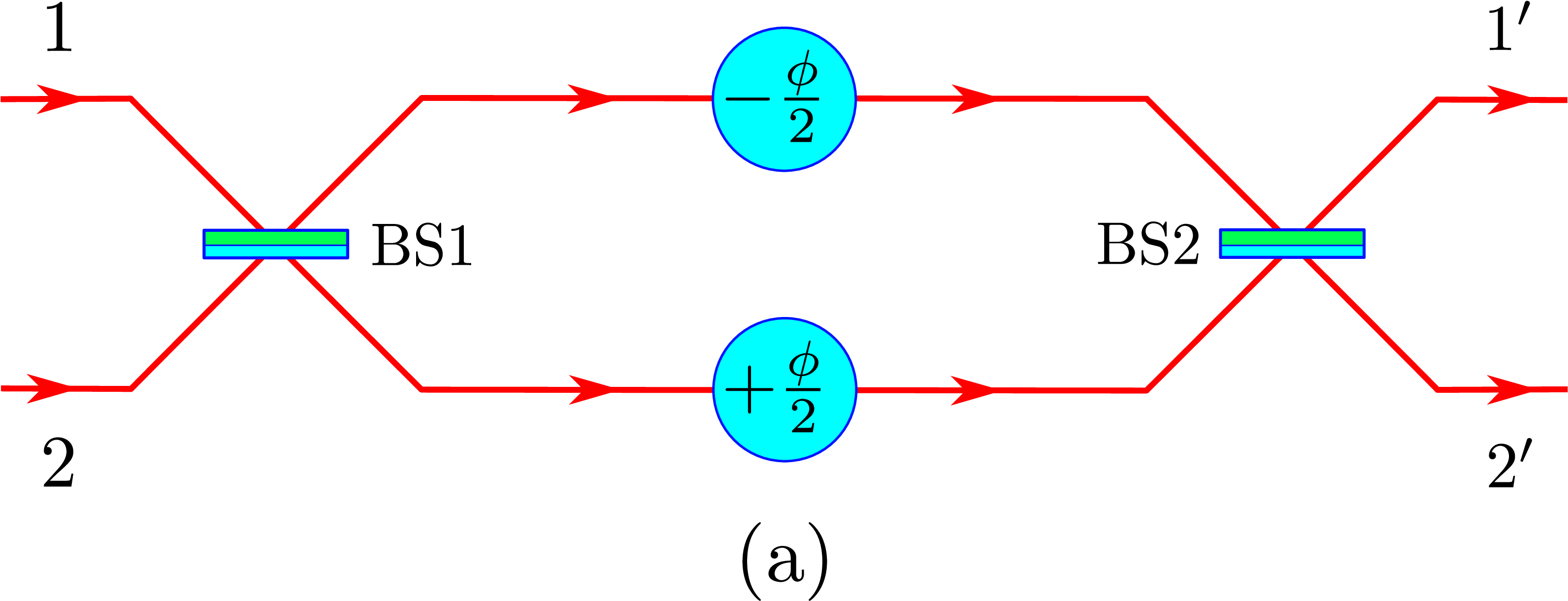}\\*[5mm]
\includegraphics*[width=160mm]{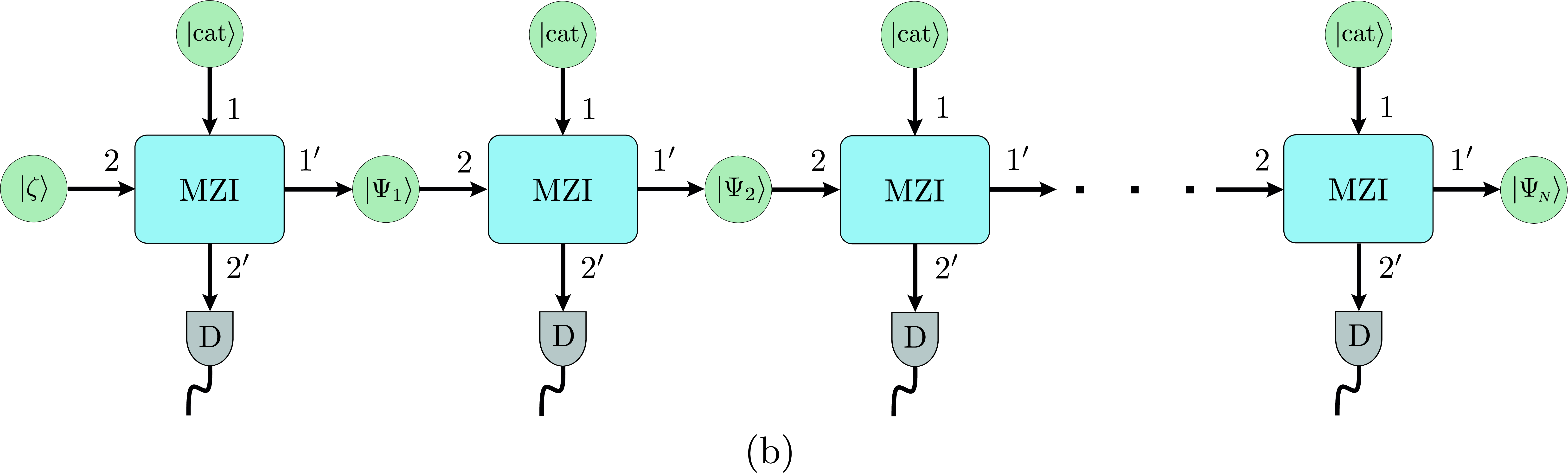}
\caption{Panel (a) shows the Mach-Zehnder interferometer (MZI) for the proposed controlled-displacement operation
key to the QW encoding.
Here BS1 and BS2 are both 50:50 beamsplitters, and the phase shifters
on the two arms induce the phase shifts $\pm \phi/2$ as indicated. The beamsplitter parameters are chosen
so that the MZI effects the input-output relations \eref{io_reln} on the input and the output mode operators.
Panel (b) illustrates the schematics for the QW encoding based on the MZI of (a). The encoding
begins from the left with a squeezed-vacuum state $|\zeta\rangle$ along with an ancillary cat state
$|{\rm cat}\rangle$ of \eref{cat} input to the MZI. Conditioned on measurement outcomes for output port $2'$ at the
detector D (see text for details), the MZI output port $1'$ would produce the one-run encoded approximate GKP
state $|\Psi_1\rangle$. Feeding $|\Psi_1\rangle$ together with another ancillary cat state into the MZI, the same
post-selection procedure as previously would yield in the output port $1'$ the next-run encoded state. Repeating
this procedure $N$ times, one would generate at the final output the $N$-run encoded state $|\Psi_N\rangle$ of \eref{Psi_N}.
\label{fig:MZI}}
\end{figure}

As pointed out above, in the QW approach to bosonic grid-code generation, the codeword pattern is determined
in accordance with the coin operation for the QW. Throughout this work, we shall adopt a two-state quantum coin
(thus a ``coin qubit"), so that each step of the QW is conditioned on the coin configurations $\{|R\rangle,|L\rangle\}$,
corresponding to rightward ($R$) and leftward ($L$) displacements by one single step length.
To prepare an approximate GKP state with a binomial profile, it is found in Ref.~\cite{Li20} that
the corresponding coin operation can be chosen
\begin{eqnarray}
\hat{\cal C} = |D\rangle\langle D| \, ,
\label{C_D}
\end{eqnarray}
where
\begin{eqnarray}
|D\rangle \equiv (|R\rangle + |L\rangle)/\sqrt{2}
\label{D_state}
\end{eqnarray}
is the ``diagonal" coin-state. The
corresponding ``walk operator" would then take the following form in the state space of
(coin)$\otimes$(bosonic mode)
\begin{eqnarray}
\hat{W} = \hat{\cal T}(\alpha_d) \left(\hat{\cal C} \otimes \hat{I}\right) \, .
\label{W_op}
\end{eqnarray}
Here $\alpha_d$ is the step length of the QW, $\hat{I}$ is the identity operator of the bosonic mode,
and $\hat{\cal T}$ is a translation operator whose action is conditioned on the coin configuration
\begin{eqnarray}
\hat{\cal T}(\alpha_d) = |R\rangle\langle R| \otimes \hat{D}(+\alpha_d)
+ |L\rangle\langle L| \otimes \hat{D}(-\alpha_d)  \, .
\label{T_op}
\end{eqnarray}
For instance, to encode the bosonic mode into an approximate GKP logical-zero state, one may
start by preparing the mode in a position-squeezed vacuum state along with a coin qubit initialized in
the $|R\rangle$ state. As demonstrated in Ref.~\cite{Li20}, after even steps of QW with step length
$\alpha_d = \sqrt{\pi/2}$ in phase space, the coin transformation \eref{C_D} would enable approximate GKP
logical-zero states to be encoded into the bosonic mode, with a \Revise{codeword state} fidelity depending on
the level of squeezing and the number of QW steps \cite{Li20}. In this work,
we propose to achieve such encoding by means of a linear-optical setup that consists mainly
of a Mach-Zehnder interferometer (MZI) as illustrated in Fig.~\ref{fig:MZI}(a). As we will show below,
under appropriate parameter regime the MZI-device will enact the targeted QW encoding when aided with
certain non-Gaussian resource state and post-selection.

\subsection{\Revise{Linear-optical realization for quantum-walk encoding}}
To begin with, let us consider inputting a coherent state $|\alpha\rangle$ into port 1 and a position-squeezed vacuum state
$|\zeta\rangle$ into port 2 of the MZI. Here the MZI parameters are chosen to induce the following input-output
relations among the mode operators of its input and output modes
\begin{eqnarray}
\left(
  \begin{array}{c}
    \hat{a}_{1'}\\
    \hat{a}_{2'}
  \end{array}
\right)
=
\left(
  \begin{array}{cc}
    \sin\frac{\phi}{2} &  \cos\frac{\phi}{2} \\
    \cos\frac{\phi}{2} & -\sin\frac{\phi}{2}
  \end{array}
\right)
\left(
  \begin{array}{c}
    \hat{a}_1\\
    \hat{a}_2
  \end{array}
\right) \, ,
\label{io_reln}
\end{eqnarray}
where the mode indices and the phase angle $\phi$ are indicated in Fig.~\ref{fig:MZI}(a).
For small phase difference $\phi$ in the MZI and weakly squeezed input state, one can show that the MZI enacts
the map \cite{note_sqz,Wu19}
\begin{eqnarray}
|\alpha\rangle_1 |\zeta\rangle_2 \xlongrightarrow{\rm MZI} \hat{D}_{1'}\!\left(\alpha_\phi\right) |\zeta\rangle_{1'}|\alpha\rangle_{2'}
\equiv |\alpha_\phi,\zeta\rangle_{1'}|\alpha\rangle_{2'} \, ,
\label{MZI_map1}
\end{eqnarray}
where $\alpha_\phi\equiv \alpha\phi/2$ and the subscripts indicate the corresponding input/output modes; thus
$\hat{D}_{1'}(\beta)$ is the displacement operator for output mode $1'$ with displacement $\beta$. Namely,
from \eref{MZI_map1}, in the limit of small $\phi$ and $|\zeta|$ the MZI would swap the coherent state $|\alpha\rangle$ from mode 1 to
mode $2'$, while mapping the squeezed state $|\zeta\rangle$ from mode 2 to a squeezed coherent state $|\alpha_\phi,\zeta\rangle$
in mode $1'$. The MZI thus executes a phase-space displacement of mode 2 by an amount dependent on the mode 1 input, which is then
followed by a SWAP operation. As we will now show, by replacing the input of mode 1
with designated state, one would be able to realize conditional displacements that are crucial to the QW encoding.

Let us now replace the input to port 1 of the MZI with the cat state
\begin{eqnarray}
|{\rm cat} \rangle \equiv \frac{1}{\sqrt{N_\alpha}} \left(|+\alpha\rangle + |-\alpha\rangle \right) \, ,
\label{cat}
\end{eqnarray}
where $N_\alpha\equiv 2 (1+e^{-2|\alpha|^2})$ is a normalization factor. In a ``coherent-state logic" approach to
quantum information processing that
treats $|\pm \alpha\rangle$ as the logical basis states \cite{Je02,Ra03}, the cat state \eref{cat} plays the role of a
``diagonal state" similar to \eref{D_state}. In the same limit of small phase difference $\phi$ as above,
the MZI would now lead to the mapping
\begin{eqnarray}
|{\rm cat}\rangle_1 |\zeta\rangle_2 \xlongrightarrow{\rm MZI}  \frac{1}{\sqrt{N_\alpha}} \left(|+\alpha_\phi,\zeta\rangle_{1'}|+\alpha\rangle_{2'} + |-\alpha_\phi,\zeta\rangle_{1'}|-\alpha\rangle_{2'} \right) \, .
\label{MZI_map2}
\end{eqnarray}
With the aid of the cat-state input, the MZI realizes here a controlled-displacement operation conditioned on the
coherent-state ``logical configurations" $|\pm \alpha\rangle$. That is, upon being transferred to mode $1'$ through the MZI,
the squeezed vacuum state $|\zeta\rangle$
would acquire phase-space displacements $\pm \alpha_\phi$ conditioned on mode $2'$ being in the $|\pm \alpha\rangle$ states.
Comparing the MZI transformation \eref{MZI_map2} with the translation operator $\hat{\cal T}$ in \eref{T_op}, we see that
mode 2 here plays the role of the bosonic mode undergoing QWs, while mode 1 serves as the ``quantum coin" in
the QW-encoding scheme. More precisely, from the perspective of the ``coherent-state logic" approach, for any given value
of $\alpha$ one can express the displacement operation involved in \eref{MZI_map1} as a ``hybrid" controlled-displacement
gate over the joint space of the ``coherent-state coin qubit" and the bosonic mode
\begin{eqnarray}
\hat{C}_D(\alpha) \equiv |+\alpha\rangle_1\langle+\alpha| \otimes \hat{D}_2(+\alpha_\phi) +
|-\alpha\rangle_1\langle -\alpha| \otimes \hat{D}_2(-\alpha_\phi) \, ,
\label{CD_eff}
\end{eqnarray}
which bears close resemblance with \eref{T_op}. Here, however, since the ``coin" basis states $\{|\pm\alpha\rangle\}$ are not
orthogonal, the gate \eref{CD_eff} serves only as an approximate realization of $\hat{\cal T}$ in \eref{T_op} when acting on
the ``coin space" of $\{|\pm\alpha\rangle\}$ with fixed $\alpha$. Nevertheless, since this non-orthogonality would be
negligible for the values of $\alpha$ that will concern us, and that, more importantly, our results below will not hinge on
identifying the MZI-based controlled displacement with the operator \eref{CD_eff}, the non-orthogonal ``coin states" here will
not be an issue for what follows.

To complete the first run of the QW encoding, we note that the output modes $1'$ and $2'$ in
\eref{MZI_map2} are entangled. In view of the fact that both $|\pm \alpha\rangle$ states have the same amplitude over
even photon-number states, i.e.,
\begin{eqnarray}
\langle 2k | \pm \alpha\rangle = e^{-\frac{|\alpha|^2}{2}}\frac{\alpha^{2k}}{\sqrt{(2k)!}}
\qquad \mbox{\rm for $k=0, 1, 2, \dots$} \, ,
\end{eqnarray}
one can thus project the output mode $2'$ from \eref{MZI_map2} to the even number states
$\{|n=2k\rangle; k=0, 1, 2, \dots\}$ and cast mode $1'$ to the state
\begin{eqnarray}
|\Psi_1 \rangle \equiv |+\alpha_\phi,\zeta\rangle + |-\alpha_\phi,\zeta\rangle \, ,
\label{Psi_1}
\end{eqnarray}
where we have left out the normalization factor and the mode index for brevity. This completes one single step of the QW and
furnishes a one-run encoding that prepares the bosonic mode in a ``squeezed cat-state",
which has two Gaussian spikes centering at $x=\pm \sqrt{2}\,\alpha_\phi$ in phase space
(mind the extra factor of $\sqrt{2}$ here, as noted after Eq.~\eref{sqrt2} above; see also \cite{note_sqrt2}). Comparing the state
\eref{Psi_1} with the ideal GKP codewords \eref{ideal_GKP}, we see that it provides a lowest-order approximation to the
GKP logical state $|1_L\rangle$ provided appropriate values of $\alpha_\phi$ and squeezing parameter $\zeta$ are chosen \cite{Li20}.
We point out that the projection of mode $2'$ onto even photon-number states can be achieved through photon-number resolving
detection, or in a more sophisticated manner, via photon-number ``parity measurements" that projects the mode onto the
cat state \eref{cat}, which is comprised only of even photon-number states \cite{Ge05,Ha06,Ha07,Pl10,Su14}. In the latter case,
since the detection of cat state at detector ``D" in Fig.~\ref{fig:MZI}(b) signals a success event for post-selection, the present encoding
mechanism can thus also be taken a protocol using the cat state \eref{cat} as a ``catalysis" for the encoding (cf.~Ref.~\cite{Ea19}).
In the following, we shall simply refer to the post-selection procedure as ``parity measurement", irrespective of how it is
carried out.

\begin{figure}
\centering
\includegraphics*[width=170mm]{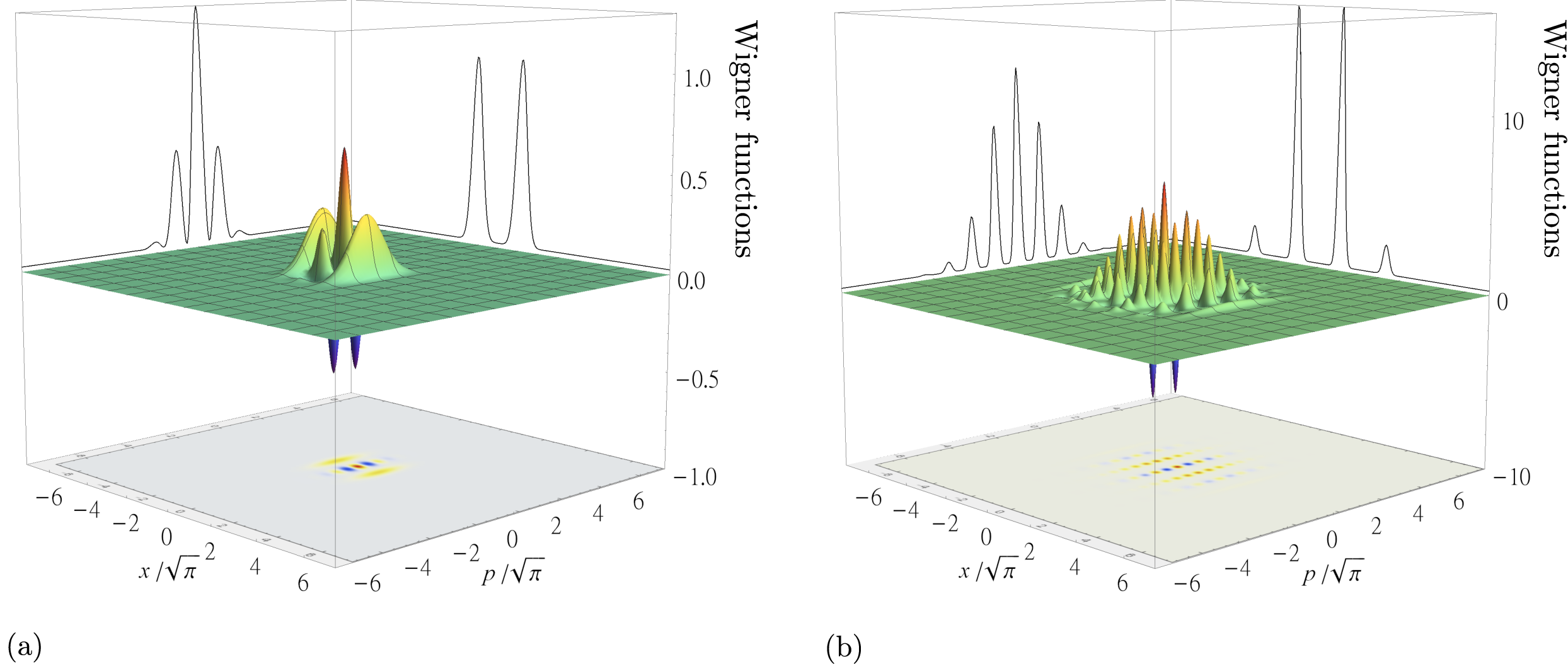}\\*[5mm]
\includegraphics*[width=170mm]{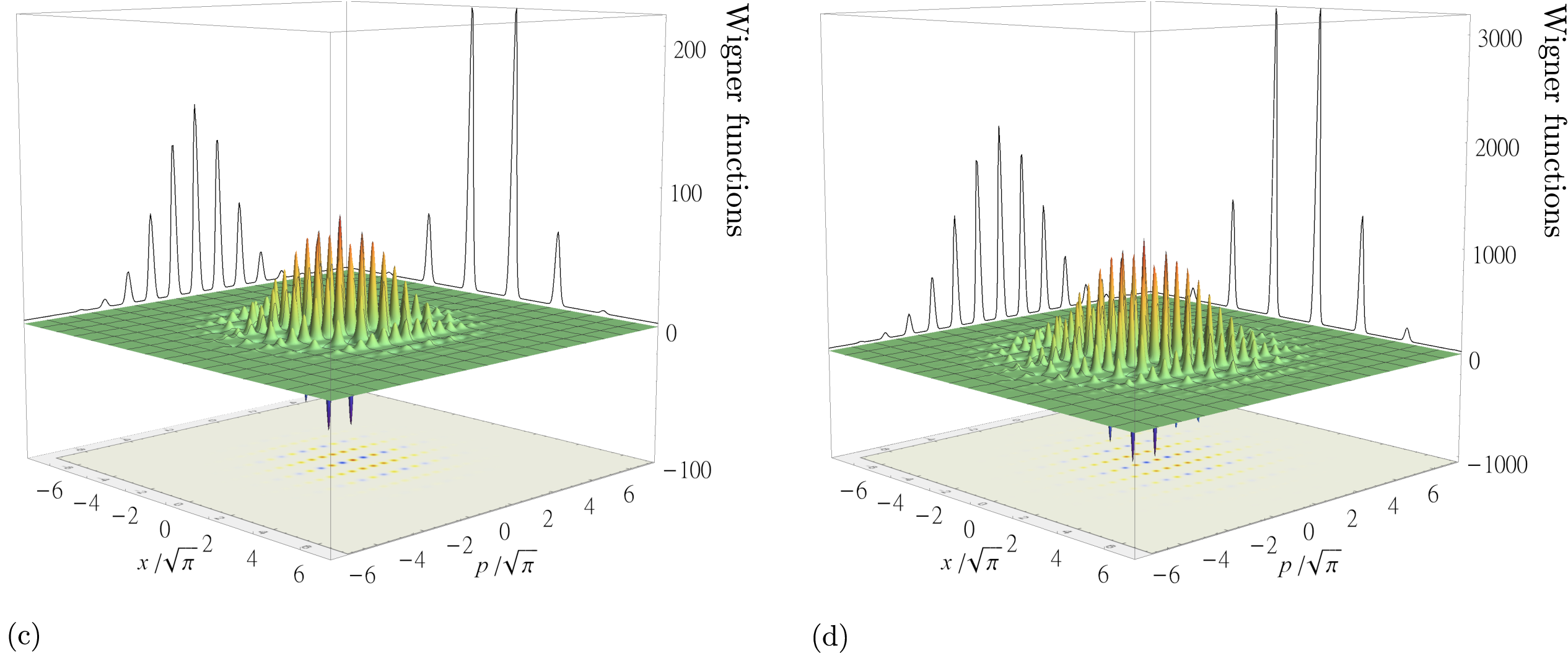}
\caption{Wigner functions for the $N$-run codeword states $|\Psi_N\rangle$ of Eq.~\eref{Psi_N} with (a) $N =1$, (b) $N=3$,
(c) $N=5$, and (d) $N=7$, where the corresponding position and momentum marginal distributions
are also shown on the side frames. Here the codeword parameters are chosen for approximating
square-lattice GKP \Revise{codeword states} (see text for details). Notice that both the $x$- and the $p$- axes are in units of
$\sqrt{\pi}$, and we have offset the marginal distribution functions slightly for clarity.
\label{fig:Wigner}}
\end{figure}

To proceed further with the QW encoding, we take the one-run encoded state \eref{Psi_1} as the new input to port 2
together with a fresh cat state \eref{cat} input to port 1 of the MZI. One can establish that the output from
the MZI would then yield
\begin{eqnarray}
|{\rm cat}\rangle_1 |\Psi_1\rangle_2 \xlongrightarrow{\rm MZI} \frac{1}{\sqrt{N_\alpha}} \,
[\,\, & (\,|+2\alpha_\phi,\zeta\rangle_{1'}+|0,\zeta\rangle_{1'}\,) \, |+\alpha\rangle_{2'}
\nonumber \\
& + (\,|0,\zeta\rangle_{1'}+|-2\alpha_\phi,\zeta\rangle_{1'}\,)\,|-\alpha\rangle_{2'} \,\,\, ] \, .
\label{MZI_map3}
\end{eqnarray}
The same ``parity measurement" over mode $2'$ as in the preceding run would then generate in mode $1'$ the two-run
encoded state
\begin{eqnarray}
|\Psi_2 \rangle \equiv |+2\alpha_\phi,\zeta\rangle + 2\,|0,\zeta\rangle + |-2\alpha_\phi,\zeta\rangle \, ,
\label{Psi_2}
\end{eqnarray}
where again we have omitted the normalization factor for simplicity. As previously, comparing \eref{Psi_2} with the ideal GKP
\Revise{codeword states} \eref{ideal_GKP}, we see that $|\Psi_2\rangle$ furnishes an approximate GKP basis state $|0_L\rangle$
for proper values of the parameters $\alpha_\phi$ and $\zeta$. In the same way, by directing the
two-run encoded state $|\Psi_2\rangle$ into port 2 and supplying another fresh cat state to port 1 of the MZI,
the same procedure as above would lead to the next run encoded state $|\Psi_3\rangle$. Repeating this
procedure for $N$ runs, \Revise{as shown in Appendix \ref{sec:Psi_N},} up to normalization factor, one can arrive at
the $N$-run QW-encoded state
\begin{eqnarray}
|\Psi_N\rangle \equiv \sum_{m=0}^N
\left(
\begin{array}{c}
N \\ m
\end{array}
\right)
|(2m-N)\alpha_\phi,\zeta\rangle \, ,
\label{Psi_N}
\end{eqnarray}
\Revise{which is consistent with what was found in Ref.~\cite{Li20}.}
As one can verify by comparing \eref{Psi_N} with \eref{ideal_GKP}, $|\Psi_N\rangle$ can be taken an approximate GKP logical state
$|0_L\rangle$ for even values of $N$, and $|1_L\rangle$ for odd values of $N$ when suitable values of
$\alpha_\phi$ and $\zeta$ are adopted \cite{Li20}. As noted below Eq.~\eref{T_op} previously, the QW-step length, thus the
parameter $\alpha_\phi=\alpha\phi/2$, appropriate for the GKP codeword states \eref{ideal_GKP} is $\alpha_d=\sqrt{\pi/2}$. For the
squeezing parameter $\zeta$, as shown in Ref.~\cite{Li20}, for large values of $N$ the phase-space distribution of the
codeword state \eref{Psi_N} would tend to approximate GKP states with Gaussian spikes of widths $\Delta_x=e^{-\zeta}$
in the position quadrature and $\Delta_p=1/\sqrt{N\pi}$ in the momentum quadrature. In order to achieve
approximate square-lattice GKP codewords that can cope with shift errors symmetric in
position and momentum quadratures, one must have $\Delta_x=\Delta_p$.
For the $N$-run QW \Revise{codeword state} \eref{Psi_N}, it is thus necessary to have a squeezing parameter such that
$e^{+\zeta}=\sqrt{N\pi}$.
\Revise{Explicitly, the Wigner function for the codeword $|\Psi_N\rangle$ can be found to be \cite{Ba97,Le97}
\begin{eqnarray}
W_N(x,p) &\equiv& \frac{1}{2\pi} \int_{-\infty}^\infty\!\! dq\,\,e^{ipq}
\left\langle\left. x-\frac{q}{2}\right|\Psi_N\right\rangle \left\langle\Psi_N\left|x+\frac{q}{2}\right.\right\rangle
\nonumber \\
&=& \frac{1}{\pi} \sum_{m=0}^N  \sum_{n=0}^N
\left(
\begin{array}{c}
N \\ m
\end{array}
\right)
\left(
\begin{array}{c}
N \\ n
\end{array}
\right) \exp\!\left\{2\,(m-n)^2\alpha_\phi^2\,e^{+2\zeta}\right\}
\nonumber\\
&\times&
\exp\!\left\{-\frac{1}{2e^{-2\zeta}}\left[(x-\sqrt{2}\,(2m-N)\alpha_\phi)^2 + (x-\sqrt{2}\,(2n-N)\alpha_\phi)^2\right]\right\}
\nonumber\\
&\times&
\exp\!\left\{-\frac{p^2}{e^{+2\zeta}}\right\} \cos\!\left(2\sqrt{2}\,(m-n)\alpha_\phi\,p\right) \, ,
\label{W_N}
\end{eqnarray}
where $|x\pm q/2\rangle$ are position basis states. Using $\alpha_\phi=\sqrt{\pi/2}$ and $e^{+\zeta}=\sqrt{N\pi}$
in \eref{W_N}}, we show in Fig.~\ref{fig:Wigner} the Wigner functions for the first few runs of
the QW \Revise{codeword states} \eref{Psi_N} that intend to approximate the GKP logical basis state $|1\rangle_L$.
We note that these
codeword states are prepared based on the MZI map \eref{MZI_map1}, which holds only approximately under the
assumption of small MZI phase $\phi$. In practice, therefore, the present linear-optical scheme can only prepare
the codeword states \eref{Psi_N} with non-ideal fidelity. It is thus
crucial to identify the parameter regimes over which the MZI operation \eref{MZI_map1} would be more reliable, and the
codeword state \eref{Psi_N} can be prepared with high fidelity. We shall now turn to this discussion.

\subsection{\Revise{Further analysis and discussions on the results}}
The key assumption behind the MZI map \eref{MZI_map1} that enables the phase-space controlled-displacement in the
QW encoding is that the phase difference $\phi$ between the two arms of the interferometer has to be small \citep{note_sqz}.
Since the MZI map \eref{MZI_map1} is the basic operation that underlies the proposed linear-optical scheme, to assess
the quality of the \Revise{codeword states} generated from this scheme, a suitable metric can be provided through
the fidelity of the MZI map \eref{MZI_map1}. Now that for given \Revise{codeword state} \eref{Psi_N}, the parameters
$\alpha$, $\phi$, and $\zeta$ would stay fixed throughout the encoding process, the corresponding MZI-map fidelity can
therefore be captured by comparing the approximate output state in \eref{MZI_map1} with reference to the corresponding
exact output state for the specific MZI input $|\alpha\rangle_1|\zeta\rangle_2$.
The details of this calculation is shown in Appendix \ref{sec:Fidelity}, and Fig.~\ref{fig:fidelity} illustrates
some typical results for the fidelity of the MZI map \eref{MZI_map1}. Anticipating the befitting
parameter regime for the MZI map, and also aiming to simplify the calculation, we focus on the regime
$|\phi|\lesssim 1$ in finding the fidelity, which allows us to arrive at an analytic expression in Eq.~\eref{fid_eq}
for the result. In Fig.~\ref{fig:fidelity}, we plot the fidelity of the MZI map versus the
coherent-state amplitude $\alpha$ and the MZI phase $\phi$ for squeezed vacuum states with squeezing levels
that correspond to different runs of the QW encoding. The red solid curve in each panel of
Fig.~\ref{fig:fidelity} depicts the sets of $\alpha$ and $\phi$
that would combine to yield phase-space displacements for preparing square-lattice GKP qubits, i.e.,
$\alpha\phi/2=\sqrt{\pi/2}$, which we shall refer to as the ``GKP line" below for convenience. Notice that in
Fig.~\ref{fig:fidelity}, since the fidelity \eref{fid_eq} is even in $\phi$, we plot the results only for positive $\phi$.
Also, since the result \eref{fid_eq} is obtained perturbatively for $|\phi|\lesssim 1$, each plot in Fig.~\ref{fig:fidelity}
covers only up to $\phi=0.8$.

\begin{figure}
\centering
\includegraphics*[width=150mm]{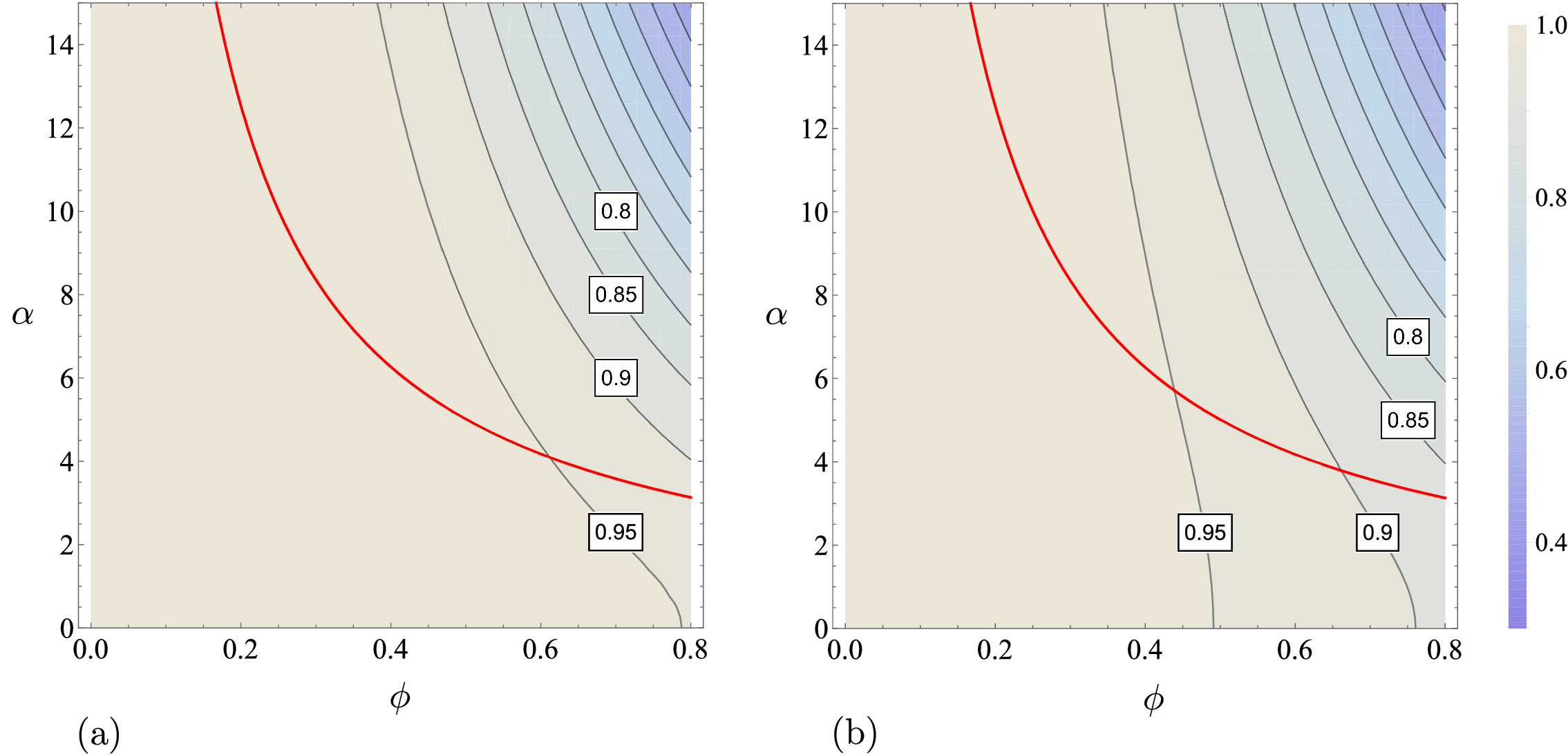}\\*[5mm]
\includegraphics*[width=150mm]{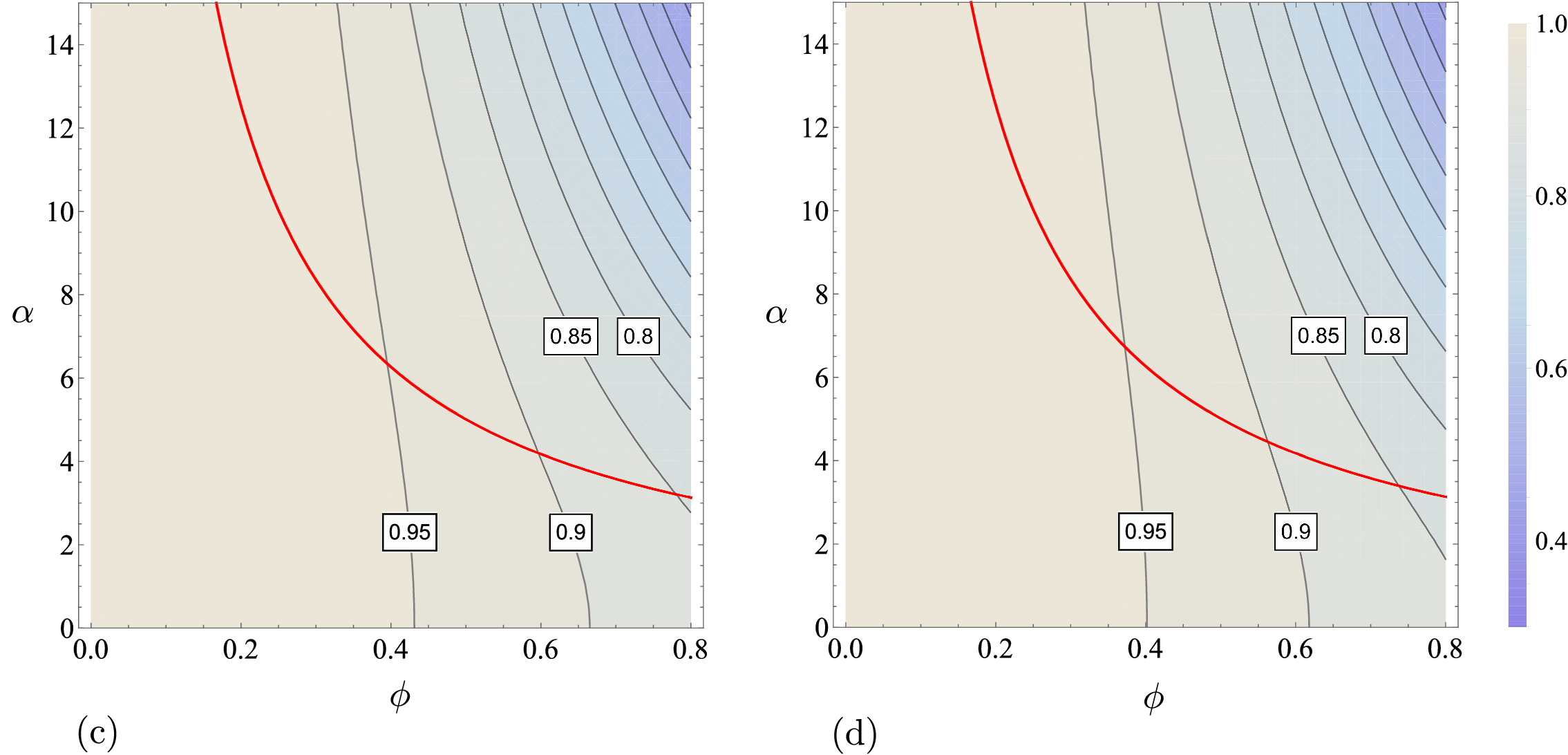}
\caption{Contour plots for the fidelity of the MZI map \eref{MZI_map1} as a function of the ancillary coherent-state amplitude $\alpha$ and the MZI phase difference $\phi$. The squeezing parameter of the squeezed vacuum state is here given by $\zeta=\frac{1}{2}\ln(N\pi)$, as per the value for an $N$-run QW encoding (see text),
with (a) $N =1$ ($\sim 4.97$ dB squeezing), (b) $N=3$ ($\sim 9.74$ dB squeezing), (c) $N=5$ ($\sim 11.96$ dB squeezing), and
(d) $N=7$ ($\sim 13.42$ dB squeezing). The red solid curve in each panel sketches the line
$\alpha\phi/2 = \sqrt{\pi/2}$, which corresponds to displacement appropriate for generating approximate square-lattice GKP \Revise{codeword states}.
The contours in each panel are spaced at 0.05 in the values of fidelity, and we have labeled
contours with fidelity 80$\%$ or higher to enhance readability. Since here the fidelity is
an even function of $\phi$, we plot only for positive $\phi$.
All plots share the same color code, which is shown next to panels (b) and (d).
\label{fig:fidelity}}
\end{figure}

\begin{figure}
\centering
\includegraphics*[width=160mm]{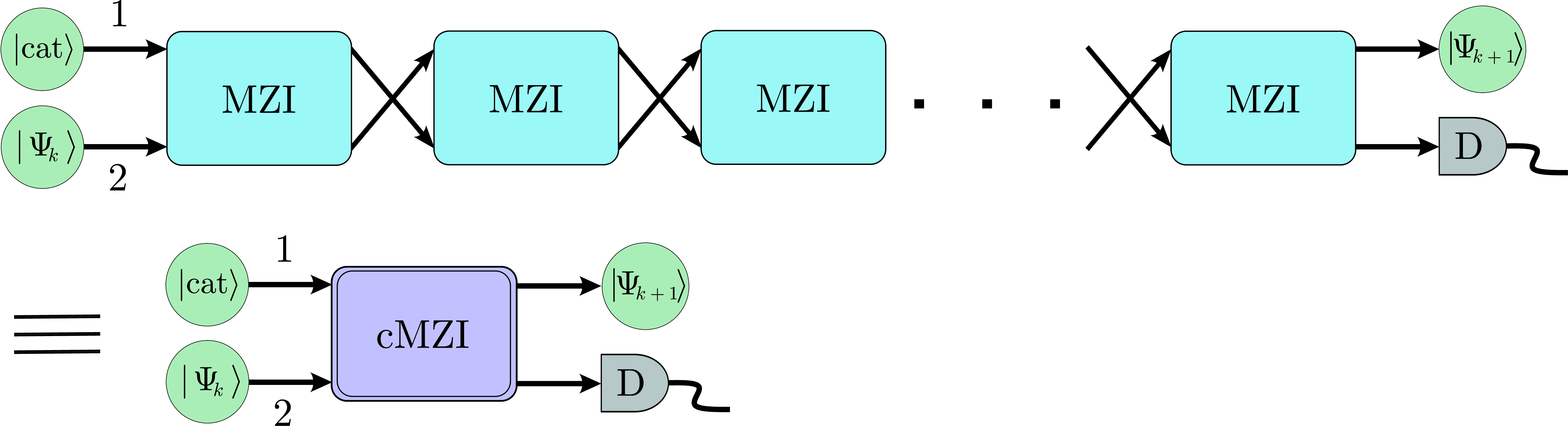}
\caption{\Revise{Concatenated MZI (cMZI) setup for enhancing phase-space displacements when only cat states of
small amplitudes are available. Notice that in the sequence of MZI operations here, since each MZI effects a swap
operation besides the phase-space displacement [see Eqs.~\eref{MZI_map1} and \eref{MZI_map2}], the output modes
from each MZI need be swapped when entering the next MZI, as indicated by the pairs of crossed arrows in the
upper illustration. After the last MZI, as before, post-selecting the ancillary mode
in even photon-number states at detector D prepares the final output mode in the desired encoded state. Here the cMZI setup
generates from the $k$-run encoded state $|\Psi_k\rangle$ the next-run encoded state $|\Psi_{k+1}\rangle$,
as illustrated also using an effective cMZI symbol in the lower panel.  }
\label{fig:cMZI}}
\end{figure}

As noted earlier, for an $N$-run QW encoding the squeezing parameter appropriate for
codewords that aim to protect against symmetric shift errors is given by $\zeta=\frac{1}{2}\ln(N\pi)$, which is the
value we use in plotting the MZI-map fidelity for each $N$ in Fig.~\ref{fig:fidelity}. It is seen
that for the lowest squeezing level $\sim 4.97$ dB [$N=1$; Fig.~\ref{fig:fidelity}(a)] the fidelity for the GKP line
(red solid curve in the plot) stays above 90$\%$ for the entire parameter regime we are covering. With increasing
squeezing levels [$N=3$, 5, 7; Fig.~\ref{fig:fidelity}(b)-(d)], however, the contours of the fidelity migrate towards the
$\alpha$-axis, and the GKP line crosses into an expanding domain of parameters with fidelity lower than 90$\%$.
This is consistent with what was noted above Eq.~\eref{MZI_map1} that the MZI map would work better for small squeezing.
However, as is clear from the results shown in Fig.~\ref{fig:fidelity}, the fidelity depends only weakly on
the squeezing parameter: as long as $\phi$ is sufficiently small, the fidelity can still stay above 95$\%$
even for the highest squeezing we are considering. While the MZI-map fidelity considered
here is not directly the ``\Revise{codeword state} fidelity", as a measure for the performance of the basic operation
underlying the present encoding scheme, it provides a means for approximately assessing the quality of the \Revise{codeword state} prepared
in the proposed scheme. Since preparing an $N$-run QW codeword state \eref{Psi_N} would require operating the MZI map $N$ times,
an estimate for the corresponding \Revise{codeword state} fidelity is therefore
\begin{eqnarray}
\left(
      \begin{array}{c}
                 \mbox{$N$-run QW \Revise{codeword state} fidelity}
      \end{array}
\right)
\sim
\left(
      \begin{array}{c}
                 \mbox{MZI-map fidelity}
      \end{array}
\right)^N \, .
\label{code_fid}
\end{eqnarray}
Thus, for instance, with a 90$\%$ MZI-map fidelity, a three-run QW codeword state prepared using the present linear-optical scheme
would have a \Revise{codeword state} fidelity $\sim (0.9)^3 \sim 73\%$. And with 99$\%$ MZI-map fidelity, a three-run QW \Revise{codeword state}
can be prepared with a \Revise{codeword state} fidelity $\sim 97\%$, which would require ancillary cat states with amplitude $\alpha\sim 10$
according to Fig.~\ref{fig:fidelity}(b). In general, since the MZI map \eref{MZI_map1} is a better approximation at small $\phi$,
for finite displacement, as is clear from Fig.~\ref{fig:fidelity}, it is favorable to have ancillary cat states
with large amplitudes. In superconducting-circuit systems, it has recently been reported that cat states with
amplitude $\alpha\sim 16$ in the microwave domain can be generated experimentally \cite{Mi23}. It is therefore
possible to prepare approximate square-lattice GKP \Revise{codeword states} utilizing
the present linear-optical QW scheme. In the optical domain, however, currently the ``large" cat states accessible to experiments
have been limited to amplitudes $\alpha\sim 2$. This poses a challenge to preparing approximate square-lattice GKP states
in the optical domain following the present prescription. Before the arrival of optical cat states with amplitudes
$\alpha\gtrsim 10$ in the laboratories, in the optical domain the proposed linear-optical scheme
can be applied to prepare, for instance, grid codes with non-square lattices that can cope with
``biased" shift errors, which have noises in one quadrature worse than the other \cite{Go01}.
Alternatively, one can also attempt to generate grid states with stabilizers generating lattices with smaller
lattice constants, thus higher lattice packing in phase space, such as hexagonal GKP codeword states \cite{Go01,Al18,No19},
provided the present linear-optical scheme can be modified to construct such \Revise{codeword states}.
\Revise{As a partial resolution to the dilemma with small-amplitude cat resource states, one can modify the encoding
procedure shown in Fig.~\ref{fig:MZI}(b) by replacing the MZI's there with concatenated MZI (cMZI) setups depicted
in Fig.~\ref{fig:cMZI}. The idea is to take advantage of the entanglement in the MZI output state \eref{MZI_map2} between the
modes of squeezed coherent states $|\pm\alpha_\phi,\zeta\rangle_{1'}$ and the ancillary states $|\pm\alpha\rangle_{2'}$. Instead of
passing on to the post-selection in the original encoding scheme, if the MZI output \eref{MZI_map2} is fed into another MZI,
with mode $1'$ entering port $2$ and mode $2'$ entering port $1$ of the MZI (i.e., ``swapping" the output modes when feeding them
into the next MZI), just like in Eq.~\eref{MZI_map1},
the respective squeezed coherent states $|\pm\alpha_\phi,\zeta\rangle$ would subsequently acquire additional phase-space displacements
$\pm\alpha_\phi$ from the second MZI operation, and the overall phase-space displacements become doubled. As illustrated in Fig.~\ref{fig:cMZI}, this procedure can be repeated to further extend the phase-space displacements until the desired value is
attained. Suppose the cMZI consists of $M$ consecutive MZI operations (which we shall refer to as the ``depth" of the cMZI),
as before, post-selecting the ancillary mode in the even photon-number state after the final MZI furnishes one single step
of QW with step length $\alpha_d = M\alpha_\phi$. It should be noted that this enhanced phase-space displacement is achieved at
the expense of reduced state fidelity out of the increased MZI operations per run for the encoding. Instead of \eref{code_fid}, the state fidelity of an $N$-run codeword would become now
\begin{eqnarray}
\left(
      \begin{array}{c}
                 \mbox{fidelity of the $N$-run QW codeword state}
                 \\
                 \mbox{using depth $M$ cMZI's}
      \end{array}
\right)
\sim
\left(
      \begin{array}{c}
                 \mbox{MZI-map fidelity}
      \end{array}
\right)^{MN} \, .
\label{code_fid_cMZI}
\end{eqnarray}
It should be noted that, alternately, by increasing the MZI phase $\phi$ by a factor of $M$,
one could also achieve the same QW step length $\alpha_d = M\alpha_\phi$
using just one single MZI. In practice, therefore, it is necessary to compare the fidelity
costs between these two approaches. Depending on the fidelity landscape of the corresponding value of $N$,
the cMZI approach may or may not be the favorable option for given value of $M$.
Before applying the cMZI approach, one must optimize the value of $M$ utilizing analysis from, say, the
landscape of the MZI-fidelity shown in Fig.~\ref{fig:fidelity}.
}

In addition to non-ideal \Revise{codeword state} fidelity, another limitation of the present encoding scheme is its non-deterministic nature.
As explicated previously, for each run of the proposed linear-optical QW encoding the parity measurement that
projects the ``coin states" $|\pm\alpha\rangle$ onto the cat state \eref{cat} can be achieved through
post-selecting even photon-number states or implementing the more sophisticated photon-number parity measurement
at the ancillary output port of the MZI \cite{Ha06,Ha07,Su14}. As one can show, whichever option is chosen for
the parity measurement the post-selection rate would always be $1/2$. Preparing the $N$-run codeword state \eref{Psi_N} through
the QW encoding scheme would, therefore, have a success rate of $1/2^N$.
This is, however, inherent in the non-unitary QW process behind the encoding scheme due to the projective
coin-operation \eref{C_D} of the QW \cite{Li20}. To overcome this difficulty, it is thus necessary to develop
unitary QW processes that can generate approximate square-lattice GKP codeword states.
Alternatively, one can also choose to incorporate adaptive corrections into the QW encoding scheme. For instance,
for each run of the QW encoding, whenever the post-selection detects an odd-parity outcome, one can ``correct"
the output mode of the MZI immediately, so that the desired \Revise{codeword state} can always be generated irrespective of
the post-selection results. Similarly, one can also attempt with a scheme that leaves the correction until
the final output by using the data of parity measurements during the encoding process \cite{We18}.
Despite being non-deterministic, nevertheless, the present QW encoding scheme should still work sufficiently well for
\Revise{codeword states} with small numbers of runs. For instance, in the recent experiments of Ref.~\cite{Si23}
the square-lattice GKP qubits in the error corrections correspond to our three-run and four-run QW codeword states.
For such states, the codeword generation rates from the proposed linear-optical scheme would be, respectively,
$1/2^3=12.5\%$ and $1/2^4=6.25\%$, which should still be realistic experimentally.
\Revise{Indeed, in the architecture of continuous-variable measurement-based quantum computing,
the leading framework for fault-tolerant quantum computing based on optical platforms \cite{Bo21},
when topological codes along with analogue error-corrections are incorporated into the computing scheme \cite{Fu17,Fu18},
the fault-tolerant squeezing threshold can be reduced to
lower than 10 dB, which corresponds to the three- and four-run encodings in the present scheme.}

\section{\label{sec:concl} Concluding remarks}

In summary, we have proposed in this paper a linear-optical scheme for preparing
approximate square-lattice GKP states through the QW encoding of Ref.~\cite{Li20}.
Based on an MZI device, the encoding scheme takes advantage of cat-state ancillary mode
to enact conditional displacements of a bosonic mode in phase space. With the aid of
post-selection on the ancillary mode, the bosonic mode initially prepared in
a squeezed vacuum state is subsequently encoded into the designated grid state
through QW in phase space.
Since the MZI-based conditional displacement works only approximately,
we analyze the fidelity of the underlying basic operation and pinpoint the adequate
parameter regimes over which high-quality grid states can be generated from
the proposed encoding scheme. Although our encoding scheme has been proposed mainly for
generating optical grid states, it is equally applicable to, for instance, preparing gird states
in the microwave domain using superconducting-circuit systems.

We point out that, strictly speaking, the present linear-optical scheme is not a direct implementation
for the QW-encoding scheme originally proposed in Ref.~\cite{Li20}. On the one hand, the ``coin qubit"
in the present scheme is played by the ``coherent-state qubit" in the space of $\{|\pm\alpha\rangle\}$
for given $\alpha$. Since the ``basis states" $\{|\pm\alpha\rangle\}$ are orthogonal only approximately,
the conditional displacement operator \eref{CD_eff} is also an approximate realization of the QW translation
operator \eref{T_op}. At the same time, in the present linear-optical scheme the ``coin-toss" operation \eref{C_D}
is replaced with offline prepared cat-state input, which serves to provide non-Gaussianity for grid-state encoding.
On the other hand, in the present scheme we did not seek implementation for the initial-state preparation of
the original QW-encoding proposal, which aims to generate encoded qubits with arbitrary logical states through
QW procedures \cite{Li20}. In fact, once encoded qubits in the logical basis states are available, one can
prepare arbitrary encoded logical qubits by applying single-qubit logical gates \cite{Go01}.

Finally, as already shown in Ref.~\cite{Li20}, by changing the coin-toss operation, the QW encoding can lead to grid states
with different phase-space structures. In the present linear-optical realization of the QW encoding, by modifying
the ancillary resource state or its sequence, it should be possible to encode the bosonic mode into different grid states,
such as hexagonal GKP states. We leave these lines of investigation for our future work.

\acknowledgments{This research is supported by the National Science and Technology Council of Taiwan through grants
MOST 111-2119-M-008-002, NSTC 111-2627-M-008-001, 112-2119-M-008-007, 113-2119-M-008-010,
and also NSTC 111-2923-M-032-002-MY5, and 112-2112-M-032-008-MY3 for JYW,
and MOST 111-2112-M-194-005, NSTC 112-2112-M-194-004 for STW. This work is also supported partly by
the National Center for Theoretical Sciences, Taiwan.}

\appendix
\section{\label{sec:Psi_N}Derivation for the $N$-run codeword state \eref{Psi_N}}
\Revise{
In this appendix, we establish the expression for the codeword state $|\Psi_N\rangle$ of \eref{Psi_N}
generated from an $N$-run QW-encoding process as illustrated in Fig.~\ref{fig:MZI}(b). We have shown in
the text that by means of the one-run codeword state \eref{Psi_1}, one can produce the two-run codeword \eref{Psi_2}
through another run of QW-encoding. To show that the formula \eref{Psi_N} holds for all $N=1, 2, 3, \dots$
for any $N$-run QW-encoded states, according to the method of induction, it suffices
to show that another run of QW-encoding of the $k$-run codeword state $|\Psi_k\rangle$ can prepare the $(k+1)$-run
codeword state $|\Psi_{k+1}\rangle$ as given by \eref{Psi_N} with $N=(k+1)$. }

\Revise{
As usual, let us consider an MZI input with the cat state \eref{cat} into port 1 and the $k$-run codeword state
$|\Psi_k\rangle$ into port 2 of the MZI. We then have here
\begin{flalign}
|{\rm cat}\rangle_1 |\Psi_k\rangle_2
& = \frac{1}{\sqrt{N_\alpha}} \left(|+\alpha\rangle_1 + |-\alpha\rangle_1 \right) \otimes
\sum_{m=0}^k \left( \begin{array}{c} k \\ m \end{array} \right)
|(2m-k)\alpha_\phi,\zeta\rangle_2  &
\nonumber \\
& = \frac{1}{\sqrt{N_\alpha}} \sum_{m=0}^k \left( \begin{array}{c} k \\ m \end{array} \right)
\left( |+\alpha\rangle_1 |(2m-k)\alpha_\phi,\zeta\rangle_2
     + |-\alpha\rangle_1 |(2m-k)\alpha_\phi,\zeta\rangle_2 \right) &
\nonumber \\
&\xlongrightarrow{\rm MZI} \frac{1}{\sqrt{N_\alpha}} \sum_{m=0}^k \left( \begin{array}{c} k \\ m \end{array} \right)
\left( \,\, |(2m-k+1)\alpha_\phi,\zeta\rangle_{1'} |+\alpha\rangle_{2'}
 + |(2m-k-1)\alpha_\phi,\zeta\rangle_{1'} |-\alpha\rangle_{2'} \, \right) \, , &
\label{MZI_mapk}
\end{flalign}
Postselecting mode $2'$ through the parity measurement detailed in the text would then yield from \eref{MZI_mapk}
for mode $1'$ (save the prefactor $1/2$ and the mode index $1'$ for brevity)
\begin{flalign}
& \sum_{m=0}^k
\left(\begin{array}{c} k \\ m \end{array} \right) |(2m-k+1)\alpha_\phi,\zeta\rangle
+ \sum_{m=0}^k
\left(\begin{array}{c} k \\ m \end{array} \right) |(2m-k-1)\alpha_\phi,\zeta\rangle &
\nonumber \\
& = \sum_{m=1}^{k+1}
\left(\begin{array}{c} k \\ m-1 \end{array} \right) |(2m-k-1)\alpha_\phi,\zeta\rangle
+ \sum_{m=0}^k
\left(\begin{array}{c} k \\ m \end{array} \right) |(2m-k-1)\alpha_\phi,\zeta\rangle &
\nonumber \\
& = |-(k+1)\alpha_\phi,\zeta\rangle + |(k+1)\alpha_\phi,\zeta\rangle
+ \sum_{m=1}^{k} \left[\left(\begin{array}{c} k \\ m-1 \end{array} \right) +
\left(\begin{array}{c} k \\ m \end{array} \right) \right]
|(2m-k-1)\alpha_\phi,\zeta\rangle &
\nonumber \\
& = \sum_{m=0}^{k+1} \left( \begin{array}{c} k+1 \\ m \end{array} \right)
|(2m-(k+1))\alpha_\phi,\zeta\rangle
= |\Psi_{k+1}\rangle \, , &
\label{Psi_kk}
\end{flalign}
where in arriving at the final expression we have used the identity
\begin{eqnarray}
\left(\begin{array}{c} k \\ m-1 \end{array} \right) +
\left(\begin{array}{c} k \\ m \end{array} \right)
=  \left( \begin{array}{c} k+1 \\ m \end{array} \right)  \, .
\end{eqnarray}
From \eref{Psi_kk}, we see that the $k$-run codeword state $|\Psi_k\rangle$ can be used to generate the
$(k+1)$-run codeword state $|\Psi_{k+1}\rangle$ through an additional run of the QW-encoding.
By mathematical induction, this proves that the general expression \eref{Psi_N} applies to any
$N$-run QW codeword state.
}

\section{\label{sec:Fidelity}Fidelity of the MZI map \eref{MZI_map1}}
We demonstrate in this appendix the calculation for the fidelity of the MZI map \eref{MZI_map1}. To this end,
we will derive first an operator-ordering theorem that will be
essential to this calculation, and then proceed to finding the MZI-map fidelity utilizing a coherent-state representation.
For the fidelity, let us consider an operation that displaces the squeezed vacuum state $|\zeta\rangle$
with the aid of an ancillary coherent state with amplitude $\alpha$ through the MZI. Starting with the MZI input
state $|\alpha\rangle|\zeta\rangle$, we have in accordance with the input-output relation \eref{io_reln}
the following output state
\begin{eqnarray}
|\Psi_{\rm out}\rangle \equiv &&\hat{D}\!\left(\alpha\sin\frac{\phi}{2},\alpha\cos\frac{\phi}{2}\right) \,
\nonumber \\
&&\exp\!\left[-\frac{\zeta}{2}
\left(\cos\frac{\phi}{2}\,\hat{a}^\dagger_1-\sin\frac{\phi}{2}\,\hat{a}^\dagger_2\right)^{\!\!2}
+\frac{\zeta^*}{2}
\left(\cos\frac{\phi}{2}\,\hat{a}_1-\sin\frac{\phi}{2}\,\hat{a}_2\right)^{\!\!2}
\right] |00\rangle \, ,
\label{psi_out}
\end{eqnarray}
where $\hat{D}(\alpha_1,\alpha_2)\equiv\hat{D}_1(\alpha_1)\otimes\hat{D}_2(\alpha_2)$ is the two-mode phase-space
displacement operator.
Notice that in \eref{psi_out} and what follows, for brevity, we omit the prime symbol over the mode indices $1$ and $2$.
In the limit of $|\phi|\ll 1$ and weak squeezing, one can show that the output state \eref{psi_out} would be approximately
\cite{Wu19}
\begin{eqnarray}
|\Psi_{\rm displ}\rangle \equiv
\hat{D}\!\left(\frac{\alpha\phi}{2},\alpha\right)
\exp\!\left(-\frac{\zeta}{2}\hat{a}^\dagger_1{}^2+\frac{\zeta^*}{2}\hat{a}_1^2\right) |00\rangle
= \hat{D}_1\!\left(\frac{\alpha\phi}{2}\right)|\zeta\rangle_1|\alpha\rangle_2 \, ,
\label{psi_displ}
\end{eqnarray}
which is the output state of the MZI map \eref{MZI_map1}. To quantify the fidelity of approximating the
actual output state \eref{psi_out} with \eref{psi_displ}, we compute the overlap between this pair of states
and define
\begin{eqnarray}
F\equiv|\langle \Psi_{\rm displ}|\Psi_{\rm out}\rangle| \, .
\label{fidelity}
\end{eqnarray}
To evaluate $F$ explicitly, it is necessary to invoke operator-ordering theorems that can reduce the
exponential operators in \eref{psi_out} into manageable form \cite{Ba97,Pu01}. Let us derive first
an operator-ordering theorem that will facilitate the fidelity calculation significantly afterwards.

\subsection{\label{sec:OpTh}Operator-ordering theorem}
For real squeezing parameter $\zeta$ in \eref{psi_out}, the operator-ordering theorem that we shall need in
evaluating the fidelity \eref{fidelity} will involve the following set of (Hermitian) operators
\begin{eqnarray}
\hat{A}_1 \equiv i\,(\hop{a}{1}^2 - \hat{a}^\dagger_1{}^2) \,, \quad \hat{A}_2 \equiv i\,(\hop{a}{2}^2 - \hat{a}^\dagger_2{}^2) \, , \quad
\hat{B} \equiv i\,(\hop{a}{1}^\dagger\hop{a}{2} - \hop{a}{1} \hop{a}{2}^\dagger) \,, \quad \hat{C} \equiv i\,(\hop{a}{1}\hop{a}{2} - \hop{a}{1}^\dagger \hop{a}{2}^\dagger) \,.
\nonumber\\
\label{ABC}
\end{eqnarray}
From $[\hat{a}_j,\hat{a}_k] =0$ and $[\hat{a}_j,\hat{a}_k^\dagger]=\delta_{jk}$ for $j,k=1,2$, one can establish the following commutation relations
among the set of operators \eref{ABC}
\begin{eqnarray}
&&[\hop{A}{1},\hop{A}{2}] = 0 \, , \quad [\hop{A}{1},\hat{B}] = 2i\,\hat{C} \, , \quad [\hop{A}{1},\hat{C}] = 2i\,\hat{B} \,,
\nonumber \\
&&[\hat{B},\hat{C}] = 2i\,\hat{A} \, , \quad [\hop{A}{2},\hat{B}] = -2i\,\hat{C} \, , \quad [\hop{A}{2},\hat{C}] = -2i\,\hat{B} \,,
\label{ABC_CR}
\end{eqnarray}
where we have denoted
\begin{eqnarray}
\hat{A} \equiv \frac{\hop{A}{1}-\hop{A}{2}}{2} \, .
\end{eqnarray}
Utilizing \eref{ABC_CR}, one can further show that we have
\begin{eqnarray}
&&[\hat{A},\hat{B}] = 2i\,\hat{C} \, , \quad [\hat{A},\hat{C}] = 2i\,\hat{B} \, , \quad [\hat{B},\hat{C}] = 2i\,\hat{A} \,,
\nonumber \\*[2mm]
&&\left[\frac{\hop{A}{1}+\hop{A}{2}}{2},\hat{A}\right] = 0 \, , \quad \left[\frac{\hop{A}{1}+\hop{A}{2}}{2},\hat{B}\right] = 0 \, , \quad
\left[\frac{\hop{A}{1}+\hop{A}{2}}{2},\hat{C}\right] = 0 \, .
\label{ABC_CR2}
\end{eqnarray}
Namely, the set of operators $\hat{A}$, $\hat{B}$, and $\hat{C}$ satisfies the SU(1,1) algebra except for factors of 2 \cite{Pu01},
and $(\hop{A}{1}+\hop{A}{2})/2$ commutes with each of these operators. This allows us to express the $\zeta$-dependent exponential operator
in \eref{psi_out} in the form (mind that $\zeta$ is real here)
\begin{eqnarray}
\exp\left( -\frac{i\zeta}{2} \frac{\hop{A}{1}+\hop{A}{2}}{2} \right) \exp\left[ -\frac{i\zeta}{2}
\left(\cos\phi\, \hat{A} - \sin\phi \, \hat{C} \right)\right]\, .
\label{exp_decomp}
\end{eqnarray}
For the second term of \eref{exp_decomp}, we shall derive the following operator-ordering theorem
\begin{eqnarray}
\hat{f}(\theta) \equiv e^{ \theta \left(\lambda_1 \hat{A} + \lambda_2 \hat{C} \right)}
= e^{p(\theta)\hat{A}}\, e^{r(\theta)\hat{C}}\,e^{q(\theta)\hat{B}} \, ,
\label{OOTh}
\end{eqnarray}
where $\lambda_1$, $\lambda_2$ are given parameters, and $p(\theta)$, $q(\theta)$, and $r(\theta)$ are unknown functions of $\theta$ that
satisfy the boundary conditions
\begin{eqnarray}
p(0) = q(0) = r(0) = 0 \, .
\label{pqr0}
\end{eqnarray}
We point out that despite $\hat{A}$, $\hat{B}$, $\hat{C}$ forming an SU(1,1)-like structure,
the operator-ordering theorems usually found for the SU(1,1) algebra \cite{Ba97,Pu01} turn out less favorable
in our case. Instead, due to the fact that we have here
\begin{equation}
\hat{B}|00\rangle = i\,(\hop{a}{1}^\dagger\hop{a}{2} - \hop{a}{1} \hop{a}{2}^\dagger) |00\rangle = 0 \, ,
\label{B00}
\end{equation}
it is the operator-ordering theorem in the form of \eref{OOTh} that will simplify our calculation significantly. Following standard procedures \cite{Ba97,Pu01}, one can show that \eref{OOTh} enforces the simultaneous differential equations among $p(\theta)$, $q(\theta)$, and $r(\theta)$
\begin{eqnarray}
\left\{ \,
\begin{array}{ccl}
r' &=& \lambda_2 \cos( 2p ) \,, \\
i q' \cos( 2r )&=& \lambda_2 \sin( 2p ) \,, \\
i q' \sin( 2r )&=& p' - \lambda_1 \,.
\end{array}
\right.
\label{pqr_DE}
\end{eqnarray}
It turns out that it is challenging to solve \eref{pqr_DE} in closed form. Since we are concerned mainly with
small values of $\phi$, it is legitimate to take $\lambda_2=\sin\phi\simeq\phi$ a small parameter in \eref{pqr_DE}
and solve for $p$, $q$, $r$ perturbatively. This yields
\begin{eqnarray}
\left\{ \,
\begin{array}{ccl}
p &=& \lambda_1 \theta + O(\lambda_2^2) \,, \\
q &=& \frac{i\lambda_2}{2\lambda_1} [\cos(2\lambda_1\theta)-1] + O(\lambda_2^2) \,,\\
r &=& \frac{\lambda_2}{2\lambda_1} \sin(2\lambda_1\theta) + O(\lambda_2^2) \,.
\end{array}
\right.
\label{pqr0}
\end{eqnarray}
Recovering the original physical parameters, i.e.,
\begin{equation}
\theta = - \frac{i\zeta}{2} \, , \quad
\lambda_1 = \cos\phi \, , \quad
\lambda_2 = \sin\phi \, ,
\end{equation}
we get from \eref{pqr0} that for small $\phi$
\begin{eqnarray}
\left\{ \,
\begin{array}{ccl}
p &\simeq& - \frac{i\zeta}{2} \cos\phi \,, \\
q &\simeq& - \frac{i}{2} \tan\phi\,[\cosh(\zeta\cos\phi)-1] \,, \\
r &\simeq& \frac{i}{2} \tan\phi\,\sinh(\zeta\cos\phi) \, .
\end{array}
\right.
\label{pqr}
\end{eqnarray}
As we shall find below, the operator-ordering theorem \eref{OOTh} along with the solutions \eref{pqr} for $p$, $q$, $r$
will allow us to calculate the MZI-map fidelity \eref{fidelity} explicitly in the small $\phi$ regime.

\subsection{\label{sec:CSrep}Coherent-state representation}
From \eref{psi_out} and \eref{psi_displ}, one can write the MZI-map fidelity \eref{fidelity} utilizing
the operator-ordering theorem \eref{OOTh}, anticipating $\langle\Psi_{\rm displ}|\Psi_{\rm out}\rangle$
to be real positive,
\begin{eqnarray}
F &=& \langle \Psi_{\rm displ}|\Psi_{\rm out}\rangle
\nonumber \\
&\simeq & \langle 00|
e^{-\frac{\zeta}{2}\left(\hop{a}{1}^2-\hop{a}{1}^\dagger{}^2\right)} \hat{D}^\dagger\left(\alpha_s,\alpha_c\right)
e^{-\frac{\zeta_c}{2}\left(\hop{a}{1}^\dagger{}^2-\hop{a}{1}^2\right)} e^{-\frac{\zeta_s}{2}\left(\hop{a}{2}^\dagger{}^2-\hop{a}{2}^2\right)}
\nonumber \\
&\times&e^{+\frac{1}{2} \tan\phi \sinh(\zeta\cos\phi)\left(\hop{a}{1}^\dagger\hop{a}{2}^\dagger-\hop{a}{1}\hop{a}{2}\right)}
e^{+\frac{1}{2} \tan\phi [\cosh(\zeta\cos\phi)-1] \left(\hop{a}{1}^\dagger\hop{a}{2}-\hop{a}{1}\hop{a}{2}^\dagger\right)}
|00 \rangle \, .
\label{f}
\end{eqnarray}
Here and in the following, we denote for brevity
\begin{eqnarray}
\alpha_c \equiv \alpha \left( 1 - \cos\frac{\phi}{2} \right)\, ,
\quad
\alpha_s \equiv \alpha \left(\frac{\phi}{2} - \sin\frac{\phi}{2} \right)\, ,
\quad
\zeta_c \equiv \zeta \cos^2\frac{\phi}{2}\, ,
\quad
\zeta_s \equiv \zeta \sin^2\frac{\phi}{2}\, .
\label{abrevs}
\end{eqnarray}
To evaluate \eref{f} explicitly, one can invoke \eref{B00} and at the same time express the single-mode and two-mode squeezing operators
in normal-ordered forms \cite{Ba97, Pu01}. One can then evaluate the resulting formula using a coherent-state representation by first
applying the completeness relation for the coherent-state basis $\{|\beta\rangle\}$ judiciously, i.e., \cite{Zi05}
\begin{eqnarray}
\int\frac{d^2\beta}{\pi} |\beta\rangle\langle\beta| = \hat{I} \, .
\label{CS_cr}
\end{eqnarray}
This calculation leads to
\begin{eqnarray}
F \simeq \sech\!\mu\,\sech\!\zeta_s \,
e^{-\frac{e^\zeta}{2}\ssech\!\zeta_s\left(\alpha_s^2\, e^{+\zeta_c}+\alpha_c^2\, e^{-\zeta_c}\right)}
\nonumber \\
\times \int\frac{d^2\beta_1}{\pi} \int\frac{d^2\beta_2}{\pi}
&& e^{-|\beta_1|^2} e^{-\frac{1}{2}\left(\tanh\!\zeta_s \, \beta_1^2 - 2\alpha_s \, e^\zeta\, \ssech\!\zeta_s \, \beta_1 \right)}
\nonumber \\
&\times& e^{\tanh\!\mu \, \beta_1^* \beta_2^*}
\nonumber \\
&\times& e^{-|\beta_2|^2} e^{+\frac{1}{2}\left(\tanh\!\zeta_s \, \beta_2^2 + 2\alpha_c \, e^\zeta \ssech\!\zeta_s \, \beta_2 \right)} \, ,
\label{CS_intg}
\end{eqnarray}
where we have denoted
\begin{eqnarray}
\mu \equiv \frac{1}{2} \tan\phi \sinh(\zeta\cos\phi)\, .
\label{mu}
\end{eqnarray}
To find the integrals in \eref{CS_intg}, it is helpful to use the following identity
\begin{eqnarray}
\int\frac{d^2\beta}{\pi} \, e^{-|\beta|^2} f(\beta)\, e^{a\beta^*} = f(a) \, ,
\label{CS_id}
\end{eqnarray}
where $a$ is a constant and $f(\beta)$ is an analytic function of $\beta$. We note that after carrying out
the $\beta_1$- (or $\beta_2$-) integral in \eref{CS_intg} utilizing \eref{CS_id}, one would have in the
remaining $\beta_2$- (likewise, $\beta_1$-) integral an exponent that contains
$\beta_2^*{}^2$ (or $\beta_1^*{}^2$), which makes the integral hard to be evaluated in closed form.
Nevertheless, one can show that such term would be of order $\phi^4$, and hence can be left out safely
for small $\phi$. The remaining expression can then be evaluated efficiently again using \eref{CS_id}.
One can thus arrive at the following expression for the MZI-map fidelity for small phase difference $\phi$
in the MZI
\begin{eqnarray}
F \simeq && \, \sech\mu \, \sech\zeta_s \,
\exp\!\left\{-\frac{e^\zeta}{2} \sech\zeta_s\, (\alpha_s^2\, e^{\zeta_c} + \alpha_c^2\, e^{-\zeta_c}) \right\}
\nonumber \\
&\times& \exp\!\left\{-\frac{\alpha_c}{2} \sech^2\zeta_c\, \tanh\mu\, (\alpha_c\, \tanh\zeta_s\,\tanh\mu -2 \alpha_s\, e^\zeta) \right\}
\, ,
\label{fid_eq}
\end{eqnarray}
which is the formula we use in preparing the plots in Fig.~\ref{fig:fidelity}.



\begin{thebibliography}{99}
\bibitem{note_sqrt2}
This extra factor of $\sqrt{2}$ can also be traced back to our convention of taking the Hermitian
part of the mode operator $\hat{a}$ to be $\hat{x}/\sqrt{2}$, which is commonly done in the GKP literatures.
\bibitem{note_sqz}
Although the condition of a ``weakly squeezed" state $|\zeta\rangle$ can be stated more
precisely by requiring that $|\zeta| \lesssim 1$, it turns out that the approximate map
\eref{MZI_map1} works fine even when $|\zeta|$ is slightly greater than 1
provided $\phi$ is sufficiently small (see the fidelity analysis later). We therefore choose to
impose the ``weakly squeezed" condition in a less stringent manner.
\end{thebibliography}
\end{document}